# Reliable Evaluation Protocol for Low-Precision Retrieval

**Kisu Yang**[1,3], **Yoonna Jang**[2], **Hwanseok Jang**[1], **Kenneth Choi**[1,4], **Isabelle Augenstein**[2†], **Heuiseok Lim**[3†]
[1]VAIV Company
[2]University of Copenhagen
[3]Korea University
[4]University of California, Berkeley

## Abstract

Lowering the numerical precision of model parameters and computations is widely adopted to improve the efficiency of retrieval systems. However, when computing relevance scores between the query and documents in low-precision, we observe *spurious ties* due to the reduced granularity. This introduces high variability in the results based on tie resolution, making the evaluation less reliable. To address this, we propose a more robust retrieval evaluation protocol designed to reduce score variation. It consists of: (1) High-Precision Scoring (HPS), which upcasts only the final scoring step to higher precision to resolve tied candidates with minimal computational cost; and (2) Tie-aware Retrieval Metric (TRM), which reports expected scores, range, and bias to quantify order uncertainty of tied candidates. Our experiments test multiple models with three scoring functions on twelve retrieval datasets to demonstrate that HPS dramatically reduces tie-induced instability, and TRM accurately recovers expected metric values. This combination enables a more consistent and reliable evaluation system for lower-precision retrieval.[1]

## 1 Introduction

Recent studies on low-precision techniques have been widely explored (e.g., quantization and compression) to enhance the efficiency and scalability of neural networks while reducing computational cost (Nagel et al.; Kurtic et al., 2024; Zhu et al., 2024; Hao et al., 2025). Without sacrificing performance, these methods aim to lower the numerical precision of model weights, gradients, and activations in training and inference, along with the retrieval stage (Choi et al., 2024; Lee et al., 2025) of retrieval-augmented generation (RAG) (Wang et al., 2024; Zhang et al., 2024, 2025). To generate informative responses, retrieving accurate candidates is crucial; otherwise, the following stages may be negatively affected and result in incoherent outputs (Chen et al., 2024b; Yadav et al., 2024; Sharma, 2025).

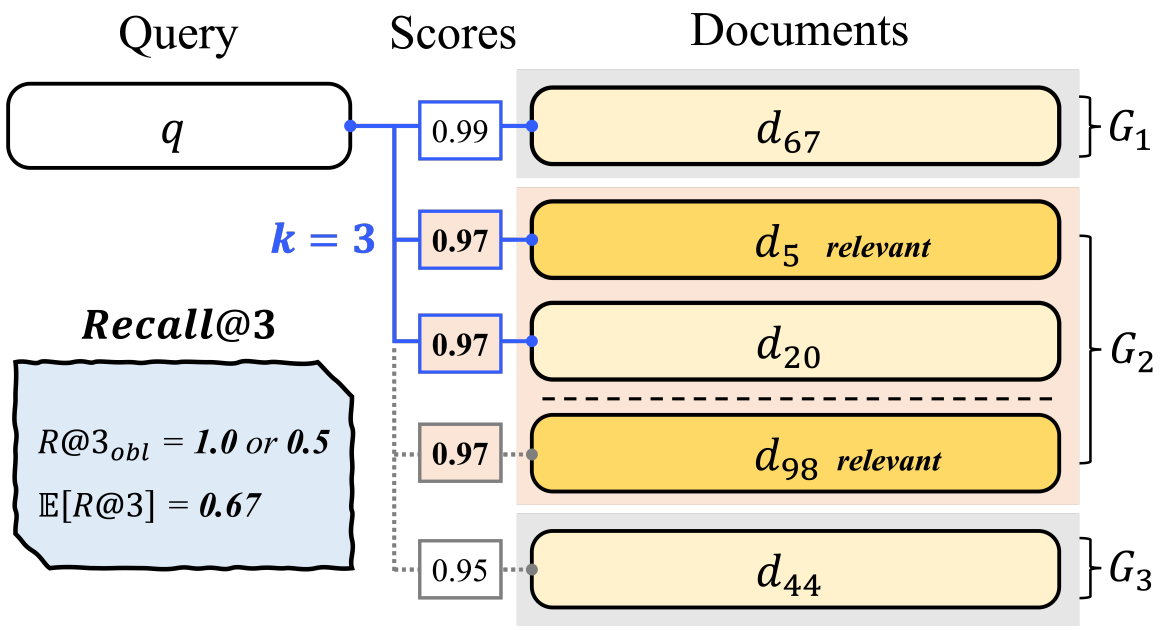

Figure 1: Example of tie-induced instability in evaluation metric. Three documents share the same score ($G_2$); two of them are relevant to the query. A tie-oblivious evaluation arbitrarily breaks the tie, so the reported R@3 depends on a random internal ordering. Instead, the tie-aware formulation deterministically reports the expectation over all permutations within the tie.

In neural retrieval systems, however, lowering numerical precision (e.g., `FP32` to `FP16`) inevitably reduces the granularity of representable floating point numbers (Shen et al., 2024; Hu et al., 2025) (see Appendix A); this coarser grid produces *spurious ties* among candidates by forcing many distinct relevance scores to quantize to the same value. Though resolving this issue can significantly affect evaluation scores (Figure 1), current mainstream retrieval evaluation systems such as MTEB[2] (Muennighoff et al., 2023) do not provide any principled mechanism for handling ties. Instead, they truncate the ranked list based on an arbitrary order (e.g., document IDs), which increases variances in results.

Thus, we propose a reliable evaluation protocol for low-precision retrieval. It is composed of (1)

†Corresponding authors.

[1]The source code is available at `https://github.com/kisuyang/tie-aware-retrieval-metrics`.

[2]`https://github.com/embeddings-benchmark/mteb`

*High-Precision Scoring* (HPS) and (2) *Tie-aware Retrieval Metric* (TRM). HPS upcasts the last scoring function into higher precision, to collapse spurious ties (Section 2.2). TRM is an expectation-based evaluation augmented with extrema (i.e., maximum and minimum achievable scores) to quantify the order uncertainty of tied candidates (Section 2.3).

Our experiments in Section 3 demonstrate that evaluating low-precision models using conventional tie-oblivious metrics leads to misleading outcomes as shown in Figure 1. Adopting HPS significantly reduces score range variability, reducing MRR@10 range by 36.82%p. Meanwhile, TRM exposes biases inherent in tie-oblivious metrics, highlighting systematic overestimation by up to +9.08%p in BF16 evaluations. By contrast, our combined approach recovers near-FP32 stability and ordering, offering a consistent and discriminative framework for evaluating retrieval models in low-precision settings.

## 2 Reliable Evaluation Protocol

We first formalize the vulnerability of the current tie-oblivious evaluation, and then present *High-Precision Scoring* and *Tie-aware Retrieval Metric*. (See Appendix A for preliminaries.)

### 2.1 Spurious Ties in Low-Precision Evaluation

Let $z$ denote the output of the linear layer after the last hidden state $h$. If the scoring function $\phi$ is softmax or sigmoid, then the cross-encoder takes the concatenated query and $i$-document pair $(q; d_i)$ as input and produces the logits $z_i$: two scalar values for softmax, or a single scalar value for sigmoid. If $\phi$ is a pairwise product, $z_i$ denotes the pair of embeddings $(h_q, h_{d_i})$ obtained by encoding the query $q$ and the document $d_i$ independently with a bi-encoder. We denote the query-document relevance score $\tilde{s}_i$ as:

$$\tilde{s}_i = \phi^{(B)}(z_i) \tag{1}$$

where $\phi^{(B)}$ indicates that $\phi$ is operated entirely in a $B$-bit mantissa format.

Applied with low-precision inference (e.g., BF16 (Burgess et al., 2019), FP16, etc), this maps theoretically continuous values onto a discrete set of representable scores; distinct true scores may collide: $\tilde{s}_i = \tilde{s}_j$ even with $z_i \neq z_j$, creating a *tie*. After sorting by $\tilde{s}$, we obtain ordered tie groups $G_n$ consisting of scores $s_i$ equivalent to $v_n$:

$$G_n = \{i \mid \tilde{s}_i = v_n\}. \tag{2}$$

```
# Forward pass
outputs = model(**inputs)
logits = outputs.logits
+ logits = logits.to(dtype=torch.float32)

# Calculate probabilities
probs = F.sigmoid(logits, dim=-1)
```

Figure 2: Implementation of HPS. We enforce FP32 precision to ensure reproducibility.

If the relevant document at cutoff rank $k$ falls inside a tie group $G_n$ where $|G_n| \geq 2$, Any evaluation that disregards ties (*tie-oblivious*) may become stochastic and yield unpredictable results, as shown in Figure 1.

### 2.2 High-Precision Scoring (HPS)

Scoring functions such as softmax, sigmoid, and pairwise product compress logits into a narrow range. This effect is exacerbated under lower-precision formats due to fewer representable values resulting in coarser bucketization in $(0, 1)$ range (see examples in Appendix B).

For lower-precision models, HPS upcasts only the final scoring operation to FP32, leaving other layers unchanged. Concretely we replace the low-precision scoring function (Equation 1) with a higher-precision scoring function:

$$\hat{s}_i = \phi(\text{upcast}(z_i)), \tag{3}$$

and retain a more fine-grained score $\hat{s}_i$ for document candidate sorting. This significantly reduces the probability of tie collisions while preserving latency, since only a small logits tensor is upcast, requiring no re-training.

**Upcast Operation.** Upcast$(\cdot)$ is a function that converts the inputs to a higher-precision floating-point datatype while preserving their real-valued magnitude. For example, a single-scalar logit of $-1.25$ that is internally represented in FP16 as the bit pattern `1 01111 0100000000` is upcast to FP32 as `1 01111111 01000000000000000000000`. Although the underlying bit representation changes, the value remains exactly $-1.25$, because every FP16 and BF16 values is exactly representable in FP32.

**Ease of Implementation.** Our proposed HPS method is designed for seamless integration into existing evaluation pipelines. As demonstrated in Figure 2, adopting HPS requires minimal modification to the codebase, often replacing a single line of code within the standard evaluation script. This drop-in compatibility ensures that researchers can reproduce our high-precision results without architectural overhaul.

**Advantages.** HPS (i) leaves the forward pass intact and upcasts logits right before scoring, (ii) adds negligible memory and time overhead as described in Appendix E, (iii) collapses large tie groups, and (iv) restores alignment with deterministic and high-precision production sorting as in Appendix C.

### 2.3 Tie-aware Retrieval Metric (TRM)

Existing tie-oblivious evaluation methods truncate the sorted list after a predefined cutoff $k$. If multiple candidates receive the same score, they are ordered arbitrarily before truncation, affecting which items are included in the top-$k$ set. As a result, the evaluation results may vary depending on how ties are resolved as illustrated in Figure 1. To mitigate this problem, TRM supplies exact *expectations*, *range*, and a *bias*.

**Expected Score.** Let $G_1, \ldots, G_N$ be tie groups sorted in descending order, where $|G_n|$ is the group size and $r_n$ is the number of relevant items. Following prior work (McSherry and Najork, 2008), we compute the expectation $\mathbb{E}[M]$ of an evaluation metric $M$ in closed form. We then use it as a diagnostic reference to quantify ordering sensitivity via the *score range* and implementation-specific deviation due to tie breaking via the *score bias*. Explicit formulas are presented in Appendix D; the linear time complexity is analyzed in Appendix E.

**Score Range.** $M_{\max}$ places the query-relevant items in each partially included tie group as early as possible; $M_{\min}$ as late as possible. For each example $i$, we report the average range over $I$ examples:

$$\mathrm{Range}(M) = \frac{1}{I}\sum_{i=1}^{I}(M_{\max,i} - M_{\min,i}). \quad (4)$$

This metric quantifies uncertainty due solely to unresolved internal orderings. A smaller range indicates that results are more stable and reliable.

**Score Bias.** Let $M_{obl,i}$ denote the tie-oblivious score for example $i$ obtained using the original implementation's fixed (typically index-preserving) ordering. We define the score bias as

$$\mathrm{Bias}(M) = \frac{1}{I}\sum_{i=1}^{I}(M_{obl,i} - \mathbb{E}[M]_i). \quad (5)$$

A large positive bias implies that $M_{obl}$ tends to overestimate the expected scores, while negative values indicate underestimation.

**Reporting Protocol.** For each cutoff $k$, we employ the $\mathbb{E}[M]$, Range($M$) and Bias($M$) as diagnostic measures to ensure evaluation reliability in this work. However, practitioners may report only the $\mathbb{E}[M]$ (or conventional $M_{obl}$) for interpretive simplicity, provided that HPS is applied or evaluation stability is internally verified via TRM.

## 3 Experiments

We evaluate to what degree our proposed evaluation protocol exposes and corrects reliability failures of existing tie-oblivious evaluation.

### 3.1 Experimental Settings

**Models.** We cover five models widely used in reranking and embedding with three prevalent scoring functions: Softmax♣, sigmoid◇, and pairwise product♠ as in Table 2.

**Datasets.** We evaluate on two primary datasets that supply a fixed candidate set per query, enabling second-stage reranker assessment independent of first-stage retrieval effects, and further extend to the MTEB-R benchmark described in Appendix F.

- **MIRACLReranking** (Zhang et al., 2023) is a multilingual reranking dataset derived from open-domain Wikipedia; its English test split contains 717 of 799 queries (excluding those without any relevant passages), each with 100 candidates (≈2.9 relevant passages on average).
- **AskUbuntuDupQuestions** (Lei et al., 2016) consists of concise questions, each with at least one annotated duplicate. The test split contains 375 queries, each with 20 candidate questions (≈6 true duplicates on average).
- **MTEB Retrieval (MTEB-R)** (Muennighoff et al., 2023) covers ten diverse English datasets: ArguAna, ClimateFEVERHardNegatives, CQADupstackGamingRetrieval, CQADupstackUnixRetrieval, FEVERHardNegatives, FiQA2018, HotpotQAHardNegatives, SCIDOCS, Touche2020Retrieval.v3, and TRECCOVID.

**Evaluation Metrics.** We evaluate the standard ranking metrics nDCG (Järvelin and Kekäläinen, 2002), MRR (Voorhees, 2000), MAP, and Recall.

| Models | FP32 | BF16 | | | | BF16 → FP32 (+HPS) | | | |
|---|---|---|---|---|---|---|---|---|---|
| | $M$ | $M_{obl}$ | $\mathbb{E}[M]$ | Range(▼) | Bias(▼) | $M_{obl}$ | $\mathbb{E}[M]$ | Range(▼) | Bias(▼) |
| **MIRACLReranking, $M$ = *nDCG*@10** | | | | | | | | | |
| Qwen3-Reranker-0.6B♣ | 73.53 | 75.04 | 68.38 | 25.59 | 6.66 | 73.59 | 73.35 | **1.13** | **0.24** |
| bge-reranker-v2-m3♢ | 74.61 | 75.59 | 74.54 | 3.90 | 1.05 | 74.63 | 74.57 | **0.16** | **0.06** |
| gte-multilingual-reranker-base♢ | 74.14 | 74.48 | 74.22 | 0.97 | 0.26 | 74.39 | 74.34 | **0.14** | **0.05** |
| Qwen3-Embedding-0.6B♠ | 63.94 | 64.52 | 63.98 | 1.90 | 0.54 | 64.01 | 64.01 | **0.00** | **0.00** |
| multilingual-e5-large♠ | 64.78 | 65.70 | 64.81 | 4.62 | 0.89 | 64.80 | 64.80 | **0.00** | **0.00** |
| **MIRACLReranking, $M$ = *MRR*@10** | | | | | | | | | |
| Qwen3-Reranker-0.6B♣ | 77.48 | 78.45 | 69.37 | 38.03 | 9.08 | 77.43 | 77.22 | **1.21** | **0.21** |
| bge-reranker-v2-m3♢ | 79.58 | 80.68 | 79.17 | 6.72 | 1.51 | 79.66 | 79.56 | **0.19** | **0.10** |
| gte-multilingual-reranker-base♢ | 79.39 | 79.75 | 79.47 | 0.85 | 0.28 | 79.59 | 79.52 | **0.18** | **0.07** |
| Qwen3-Embedding-0.6B♠ | 68.97 | 69.54 | 68.91 | 2.23 | 0.63 | 69.02 | 69.02 | **0.00** | **0.00** |
| multilingual-e5-large♠ | 71.37 | 71.84 | 71.28 | 4.61 | 0.56 | 71.18 | 71.18 | **0.00** | **0.00** |
| **AskUbuntuDupQuestions, $M$ = *MAP*@3** | | | | | | | | | |
| Qwen3-Reranker-0.6B♣ | 31.20 | 33.28 | 31.13 | 4.03 | 2.15 | 31.58 | 31.29 | **0.57** | **0.29** |
| bge-reranker-v2-m3♢ | 31.91 | 32.26 | 31.83 | 0.83 | 0.43 | 31.89 | 31.84 | **0.09** | **0.05** |
| gte-multilingual-reranker-base♢ | 30.83 | 31.23 | 30.75 | 0.93 | 0.48 | 30.69 | 30.67 | **0.03** | **0.02** |
| Qwen3-Embedding-0.6B♠ | 29.54 | 30.10 | 29.65 | 0.87 | 0.45 | 29.69 | 29.69 | **0.00** | **0.00** |
| multilingual-e5-large♠ | 29.13 | 31.31 | 29.47 | 3.54 | 1.84 | 29.70 | 29.70 | **0.00** | **0.00** |

Table 1: Results using metric $M$ with its tie-oblivious version ($M_{obl}$), expectation ($\mathbb{E}[M]$), range ($M_{\max} - M_{\min}$), and bias ($M - \mathbb{E}[M]$) on MIRACLEReranking (nDCG@10 and MRR@10) and AskUbuntuDupQuestions (MAP@3) under three precision regimes, full FP32, BF16, and BF16→FP32 (with High-Precision Scoring). In full FP32 we empirically observe $M_{obl} = \mathbb{E}[M]$ with zero range and bias, so only $M$ is shown. ♣, ♢, and ♠ indicate softmax, sigmoid, and pairwise product, respectively. Lower range and |bias| scores represent better stability.

| Models | $\phi$ | Size |
|---|---|---|
| Qwen3-Reranker-0.6B (Zhang et al., 2025) | Softmax♣ | 596M |
| bge-reranker-v2-m3 (Chen et al., 2024a) | Sigmoid♢ | 568M |
| gte-multilingual-reranker-base (Zhang et al., 2024) | Sigmoid♢ | 306M |
| Qwen3-Embedding-0.6B (Zhang et al., 2025) | Product♠ | 596M |
| multilingual-e5-large (Wang et al., 2024) | Product♠ | 560M |

Table 2: Models used in our experiments and their corresponding scoring function and size.

**Implementation Details.** We use a maximum input length of 4,096 tokens[3] and a batch size of 16[4]. All models are run under three data types: BF16, FP16, and FP32. HPS is implemented by upcasting the final scoring operation to FP32. Baseline tie-oblivious scores rely on the framework's predefined index order inside ties. In contrast, tie-aware expectations and extrema are computed with TRM.

## 3.2 Results

**Spurious Ties in Low-Precision Evaluation.** When using full BF16, the results display significant uncertainty as shown in Table 1. Especially, Qwen3-Reranker model with softmax♣ shows the highest variation — 25.59%p in nDCG@10 and 38.03%p in MRR@10. These ranges exceed the margins typically used to distinguish model superiority.

Crucially, a striking decision error appears. Under the BF16 and nDCG@$10_{obl}$ evaluation, Qwen3-Reranker seems to beat gte (75.04 > 74.48). However, tie-aware metric $\mathbb{E}$[nDCG@10] flips the ranking (68.38 < 74.22), and our proposed protocol (HPS + TRM) confirms the reversal (73.35 < 74.34) within a narrow range, rendering the naive evaluation rankings unreliable.

Albeit bias can be positive or negative, all BF16 biases are positive, implying that tie-oblivious $M_{obl}$ is overestimated (up to +9.08%p). This positive trend is likely a result of oversights in dataset construction, coupled with heuristic tie-breaking, as the positive items are more consistently placed earlier in the dataset, causing the tie-breaking to systematically favor them. We present the full experimental results on the both datasets in Figure 5 and 6.

**High-Precision with Low-Cost.** High precision scoring (HPS) collapses the large tie groups while keeping the bulk of computation in BF16. Softmax♣ ranges shrink from 25.59 to 1.13%p in nDCG@10 and from 38.03 to 1.21%p at MRR@10; sigmoid♢ model ranges drop roughly an order of magnitude (e.g., 3.90 to 0.16%p in nDCG@10); pairwise product♠ models become perfectly deterministic (range = bias = 0). The remaining softmax resid-ual range ($\sim$ 1%p) lies within ordinary inter-model differences, making rank reversals highly unlikely.

[3]Only multilingual-e5-large is truncated to 512 tokens due to its length constraints.

[4]Batch size affects the representations produced by low-precision inference, even with identical inputs.

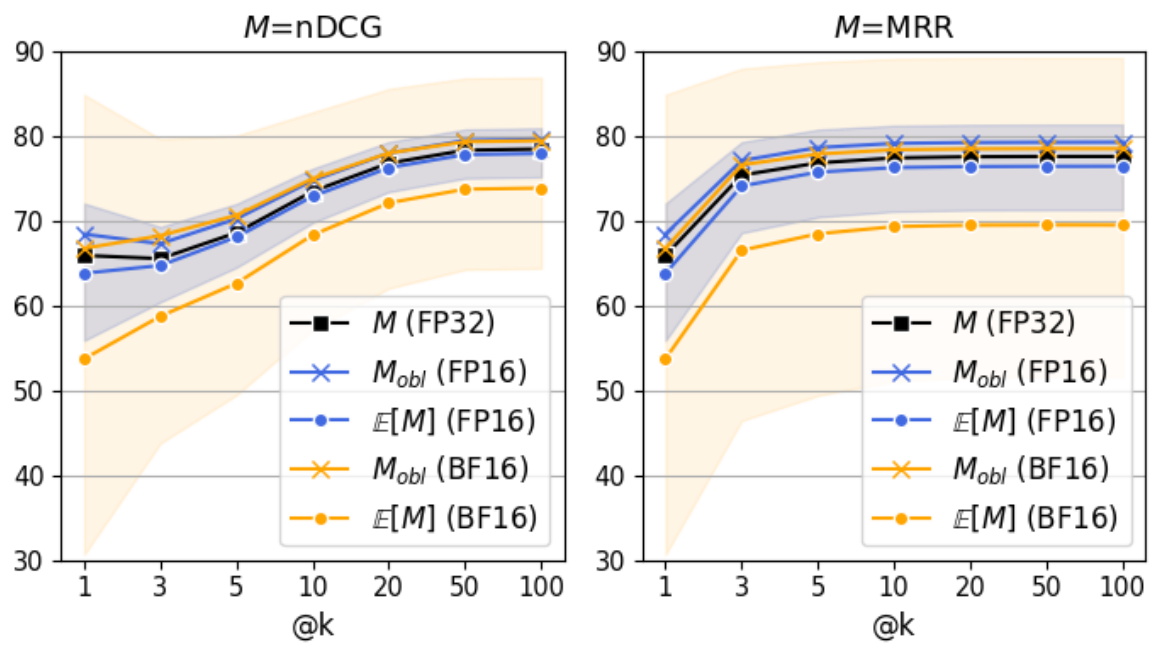

Figure 3: Tie-oblivious and expectation scores of nDCG and MRR at $k$ of `Qwen3-Reranker-0.6B`♣ model when scored with each **dtype** on MIRACLReranking.[5]

Compared to full `FP32` inference (stable but computationally costlier), HPS recovers near-`FP32` stability and ordering with negligible time and space overhead, as described in Appendix E. Consequently, while pure low-precision scoring erodes evaluation reliability, adopting our protocol, HPS with reporting ($\mathbb{E}[M]$, Range), restores precise and discriminative comparisons. We further demonstrate the superiority of this protocol over alternative baselines in Appendix G.

**Impact of Precision across Cutoffs.** Figure 3 shows nDCG and MRR metrics across various $k$-rank cutoffs, illustrating increased variance ranges and biases under lower-precision computations. Consistent with our observations in Appendix A, the `BF16` inference displays significant fluctuations and uncertainty (wide shaded areas), whereas `FP16` demonstrates intermediate stability, and `FP32` offers empirically stable results with negligible ties. This reflects the coarser bucketization induced by fewer mantissa bits in lower-precision formats (`BF16` $\ll$ `FP16` $\ll$ `FP32`).

Notably, under $M_{obl}$, the `BF16` curves surpass the `FP32` baseline at every cutoff. Such results would incorrectly indicate better performance, highlighting the unreliability of tie-oblivious evaluation due to reduced precision. Conversely, the tie-aware expectation $\mathbb{E}[M]$ consistently places `BF16` below `FP32`, accurately reflecting the true model performance.

## 4 Conclusion

We are the first to identify and formally analyze that current retrieval evaluations under low-precision settings overlook tied candidates, resulting in unstable outcomes. To address this, we propose two simple yet effective remedies: High-Precision Scoring (HPS) and Tie-Aware Retrieval Metric (TRM).

HPS upcasts the final scoring operation to collapse spurious ties with negligible computational cost, and TRM reports the expected value of evaluation scores across all possible orderings of tied candidates with score range and bias. Both methods are lightweight and straightforward to implement, and are broadly applicable across precision formats and retrieval benchmarks. Our proposed protocol mitigates spurious ties and provides a more reliable alternative to conventional evaluation practices.

As retrieval quality is a well-established prerequisite for downstream generation performance in RAG pipelines, we believe our protocol lays a more reliable foundation for evaluating and developing retrieval systems that ultimately benefit retrieval-augmented applications.

## Limitations

Our remedy targets the inference stage and does not explore how low-precision training influences ranking stability, nor whether mixed-precision training combined with HPS inference yields further gains.

We do not empirically measure the downstream impact on tasks such as retrieval-augmented generation, given that the positive correlation between retrieval performance and RAG quality is both a theoretically established premise and empirically supported by prior literature.

## Acknowledgments

This work was partly supported by the Institute of Information & Communications Technology Planning & Evaluation (IITP) grant funded by the Korean government (MSIT) (No. IITP-2026-RS-2024-00397085, Leading Generative AI Human Resources Development, 50%)

This research was co-funded by the European Union (ERC, ExplainYourself, 101077481), by the Pioneer Centre for AI, DNRF grant number P1, as well as by The Villum Synergy Programme. Views and opinions expressed are however those of the author(s) only and do not necessarily reflect those of the European Union or the European Research Council. Neither the European Union nor the granting authority can be held responsible for them.

[5] nDCG and MRR are not interval scales and thus require nuanced interpretation.

## References

Robert B Ash and Catherine A Doléans-Dade. 2000. *Probability and Measure Theory*. Academic Press.

Neil Burgess, Jelena Milanovic, Nigel Stephens, Konstantinos Monachopoulos, and David Mansell. 2019. Bfloat16 Processing for Neural Networks. In *2019 IEEE 26th Symposium on Computer Arithmetic (ARITH)*, pages 88–91. IEEE.

Jianlv Chen, Shitao Xiao, Peitian Zhang, Kun Luo, Defu Lian, and Zheng Liu. 2024a. BGE M3-Embedding: Multi-Lingual, Multi-Functionality, Multi-Granularity Text Embeddings Through Self-Knowledge Distillation.

Weijie Chen, Ting Bai, Jinbo Su, Jian Luan, Wei Liu, and Chuan Shi. 2024b. KG-Retriever: Efficient Knowledge Indexing for Retrieval-Augmented Large Language Models. *CoRR*.

Chanyeol Choi, Junseong Kim, Seolhwa Lee, Jihoon Kwon, Sangmo Gu, Yejin Kim, Minkyung Cho, and Jy-yong Sohn. 2024. Linq-Embed-Mistral: Elevating Text Retrieval with Improved GPT Data Through Task-Specific Control and Quality Refinement. Linq AI Research Blog.

David Goldberg. 1991. What Every Computer Scientist Should Know about Floating-Point Arithmetic. *ACM Computing Surveys (CSUR)*, 23(1):5–48.

Chuan Guo, Geoff Pleiss, Yu Sun, and Kilian Q Weinberger. 2017. On Calibration of Modern Neural Networks. In *International Conference on Machine Learning*, pages 1321–1330. PMLR.

Zhiwei Hao, Jianyuan Guo, Li Shen, Yong Luo, Han Hu, Guoxia Wang, Dianhai Yu, Yonggang Wen, and Dacheng Tao. 2025. Low-Precision Training of Large Language Models: Methods, Challenges, and Opportunities. *arXiv preprint arXiv:2505.01043*.

Weiming Hu, Haoyan Zhang, Cong Guo, Yu Feng, Renyang Guan, Zhendong Hua, Zihan Liu, Yue Guan, Minyi Guo, and Jingwen Leng. 2025. M-ANT: Efficient Low-bit Group Quantization for LLMs via Mathematically Adaptive Numerical Type. In *2025 IEEE International Symposium on High Performance Computer Architecture (HPCA)*, pages 1112–1126. IEEE.

Kalervo Järvelin and Jaana Kekäläinen. 2002. Cumulated Gain-based Evaluation of IR Techniques. *ACM Transactions on Information Systems (TOIS)*, 20(4):422–446.

Eldar Kurtic, Alexandre Marques, Shubhra Pandit, Mark Kurtz, and Dan Alistarh. 2024. "Give Me BF16 or Give Me Death"? Accuracy-Performance Trade-Offs in LLM Quantization. *arXiv preprint arXiv:2411.02355*.

Jinhyuk Lee, Feiyang Chen, Sahil Dua, Daniel Cer, Madhuri Shanbhogue, Iftekhar Naim, Gustavo Hernández Ábrego, Zhe Li, Kaifeng Chen, Henrique Schechter Vera, and 1 others. 2025. Gemini Embedding: Generalizable Embeddings from Gemini. *arXiv preprint arXiv:2503.07891*.

Tao Lei, Hrishikesh Joshi, Regina Barzilay, Tommi Jaakkola, Kateryna Tymoshenko, Alessandro Moschitti, and Lluís Màrquez. 2016. Semi-supervised Question Retrieval with Gated Convolutions. In *Proceedings of the 2016 Conference of the North American Chapter of the Association for Computational Linguistics: Human Language Technologies*, pages 1279–1289.

Frank McSherry and Marc Najork. 2008. Computing Information Retrieval Performance Measures Efficiently in the Presence of Tied Scores. In *European conference on information retrieval*, pages 414–421. Springer.

Paulius Micikevicius, Sharan Narang, Jonah Alben, Gregory Diamos, Erich Elsen, David Garcia, Boris Ginsburg, Michael Houston, Oleksii Kuchaiev, Ganesh Venkatesh, and 1 others. 2017. Mixed Precision Training. *arXiv preprint arXiv:1710.03740*.

Niklas Muennighoff, Nouamane Tazi, Loic Magne, and Nils Reimers. 2023. MTEB: Massive Text Embedding Benchmark. In *Proceedings of the 17th Conference of the European Chapter of the Association for Computational Linguistics*, pages 2014–2037.

Markus Nagel, Marios Fournarakis, Rana Ali Amjad, Yelysei Bondarenko, Mart van Baalen, and Tijmen Blankevoort. A White Paper on Neural Network Quantization. *arXiv preprint arXiv:2106.08295*, 4.

Chaitanya Sharma. 2025. Retrieval-Augmented Generation: A Comprehensive Survey of Architectures, Enhancements, and Robustness Frontiers. *arXiv preprint arXiv:2506.00054*.

Haihao Shen, Naveen Mellempudi, Xin He, Qun Gao, Chang Wang, and Mengni Wang. 2024. Efficient Post-training Quantization with FP8 Formats. *Proceedings of Machine Learning and Systems*, 6:483–498.

EM Voorhees. 2000. The TREC-8 Question Answering Track Report. In *Proc. Eighth Text REtrieval Conference (TREC-8), NIST Special Publication 500-246*, pages 77–82.

Liang Wang, Nan Yang, Xiaolong Huang, Linjun Yang, Rangan Majumder, and Furu Wei. 2024. Multilingual E5 Text Embeddings: A Technical Report. *arXiv preprint arXiv:2402.05672*.

Wikipedia contributors. 2025. Bfloat16 Floating-point Format.

Neemesh Yadav, Sarah Masud, Vikram Goyal, Md Shad Akhtar, and Tanmoy Chakraborty. 2024. Tox-BART: Leveraging Toxicity Attributes for Explanation Generation of Implicit Hate Speech. In *ACL (Findings)*.

Xin Zhang, Yanzhao Zhang, Dingkun Long, Wen Xie, Ziqi Dai, Jialong Tang, Huan Lin, Baosong Yang, Pengjun Xie, Fei Huang, Meishan Zhang, Li Wenjie, and Min Zhang. 2024. mGTE: Generalized Long-Context Text Representation and Reranking Models for Multilingual Text Retrieval. In *Proceedings of the 2024 Conference on Empirical Methods in Natural Language Processing: Industry Track*, pages 1393–1412.

Xinyu Zhang, Nandan Thakur, Odunayo Ogundepo, Ehsan Kamalloo, David Alfonso-Hermelo, Xiaoguang Li, Qun Liu, Mehdi Rezagholizadeh, and Jimmy Lin. 2023. MIRACL: A Multilingual Retrieval Dataset Covering 18 Diverse Languages. *Transactions of the Association for Computational Linguistics*, 11:1114–1131.

Yanzhao Zhang, Mingxin Li, Dingkun Long, Xin Zhang, Huan Lin, Baosong Yang, Pengjun Xie, An Yang, Dayiheng Liu, Junyang Lin, Fei Huang, and Jingren Zhou. 2025. Qwen3 Embedding: Advancing Text Embedding and Reranking Through Foundation Models. *arXiv preprint arXiv:2506.05176*.

Xunyu Zhu, Jian Li, Yong Liu, Can Ma, and Weiping Wang. 2024. A Survey on Model Compression for Large Language Models. *Transactions of the Association for Computational Linguistics*, 12:1556–1577.

# A Preliminaries

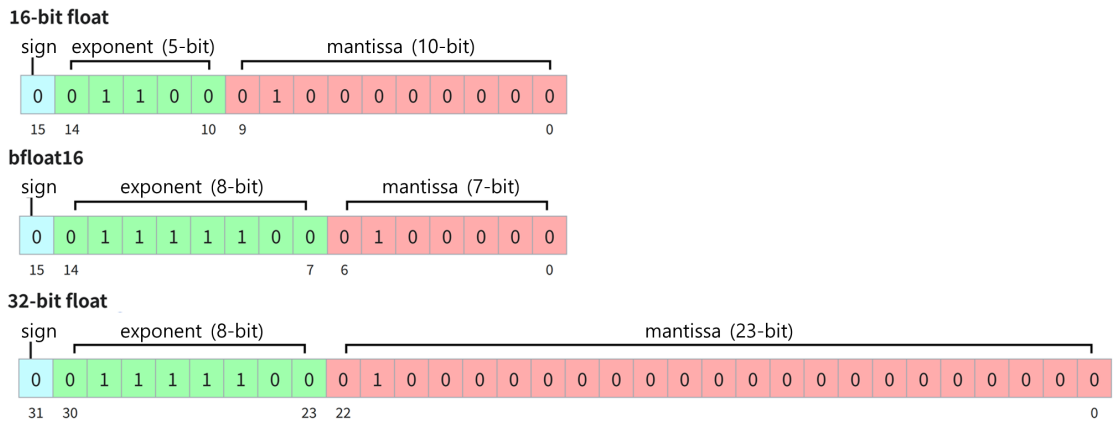

Figure 4: Bit layouts of FP16, BF16, and FP32 formats (Wikipedia contributors, 2025)

**Floating-Point Value.** A floating-point value is a way to represent numbers in computer systems, and typically encoded as three fields—*sign*, *exponent*, and *mantissa* (also called the fraction)—as illustrated in Figure 4. The exponent determines the dynamic range, the largest and smallest magnitudes that can be represented, whereas the mantissa governs the *precision* attainable within that range. Since a shorter mantissa implies coarser quantization, multiple real numbers inevitably collapse into the same representable bin, producing tied values.

After the common 1-bit sign, FP16 allocates 5 exponent bits and 10 mantissa bits, BF16 uses 8 and 7 bits respectively, and FP32 retains 8 exponent bits alongside a much longer 23-bit mantissa. By preserving the full 8-bit exponent of FP32, BF16 inherits the same dynamic range as single precision, which is widely credited with stabilizing training and thereby aiding generalization.

However, when outputs are confined to the range $(0, 1)$—as with the probabilities emitted by softmax or sigmoid scoring functions—the short 7-bit mantissa of BF16, and to a lesser extent the 10-bit mantissa of FP16, sharply reduces resolution. This loss of granularity, particularly severe in BF16, exacerbates the tied-score phenomenon and makes it difficult to distinguish among retrieval candidates that quantize to identical values.

# B Examples of Relevance Scores

The example lists below show raw relevance scores for the first query of the MIRACLReranking test split produced by the Qwen3-Reranker-0.6B model where relevant values for the given are in **bold**. The first list (scores_bf16) is obtained with both the model and scoring function executed entirely in BF16, while the second (scores_hps) applies High Precision Scoring (HPS). Tie group sizes shrink considerably under HPS.

scores_bf16 = [
1.00000000, 1.00000000, 1.00000000, 1.00000000, 1.00000000,
1.00000000, 1.00000000, 1.00000000, 1.00000000, 1.00000000,
0.99609375, 0.99609375, **0.99609375**, 0.99609375, 0.99609375,
0.99609375, 0.99609375, 0.99609375, 0.99609375, 0.99609375,
0.99609375, 0.99609375, 0.99609375, 0.99609375, 0.99609375,
0.99609375, 0.99609375, 0.99609375, 0.99609375, 0.99218750,
0.99218750, 0.99218750, 0.99218750, 0.99218750, 0.99218750,
0.99218750, 0.99218750, 0.99218750, 0.99218750, 0.99218750,
0.99218750, 0.99218750, 0.98828125, 0.98828125, **0.98828125**,
0.98828125, 0.98828125, 0.98828125, 0.98828125, 0.98437500,
0.98437500, 0.98046875, 0.97656250, 0.97656250, 0.97265625,
0.96875000, 0.96875000, 0.96875000, 0.96875000, 0.96875000,
0.96875000, 0.96093750, 0.96093750, 0.96093750, 0.96093750,
0.95703125, 0.95703125, 0.95703125, 0.95703125, 0.95703125,
0.95312500, 0.95312500, 0.94921875, 0.94921875, 0.94531250,
0.94531250, 0.94531250, 0.94140625, 0.94140625, 0.93359375,
0.92578125, 0.92578125, 0.91796875, 0.91406250, 0.88671875,
0.88671875, 0.87890625, 0.87890625, 0.87500000, 0.86718750,
0.77734375, 0.60937500, 0.51562500, 0.46875000, 0.34960938,
0.30664062, 0.28125000, 0.17285156, 0.08496094, 0.02441406,
]

scores_hps = [
0.99948066, 0.99933332, 0.99929035, 0.99919587, 0.99914408,
0.99883050, 0.99883050, 0.99875510, 0.99858958, 0.99829930,
0.99767691, **0.99767691**, 0.99752742, 0.99752742, 0.99736834,
0.99719906, 0.99719906, 0.99701905, 0.99682730, 0.99662340,
0.99662340, 0.99592990, 0.99566853, 0.99566853, 0.99509466,
0.99509466, 0.99477994, 0.99444515, 0.99444515, 0.99408901,
0.99408901, 0.99408901, 0.99330717, 0.99330717, 0.99330717,
0.99242276, 0.99142247, 0.99142247, 0.99142247, 0.99087441,
0.99087441, 0.99029154, 0.98967183, **0.98901308**, 0.98901308,
0.98901308, 0.98831278, 0.98831278, 0.98756832, 0.98593640,
0.98409361, 0.98201376, 0.97838473, 0.97702265, 0.97404259,
0.97068775, 0.96885622, 0.96885622, 0.96885622, 0.96691406,
0.96691406, 0.96267307, 0.96036118, 0.96036118, 0.96036118,
0.95791227, 0.95791227, 0.95791227, 0.95791227, 0.95531917,
0.95257413, 0.95257413, 0.94966936, 0.94966936, 0.94659668,
0.94659668, 0.94659668, 0.93991333, 0.93991333, 0.93245327,
0.92414182, 0.92414182, 0.91964257, 0.91490096, 0.88720459,
0.88720459, 0.88079703, 0.88079703, 0.87407720, 0.86703575,
0.77729988, 0.60766321, 0.51561993, 0.46879065, 0.34864515,
0.30735803, 0.28140560, 0.17328820, 0.08509904, 0.02442309,
]

# C Quantitative Analysis on Spurious Ties

We further investigate the frequency and impact of spurious ties on the MTEB-R benchmark using the Qwen3-0.6B-Reranker. By comparing various precision formats, we demonstrate how HPS restores ranking granularity and fidelity.

### C.1 Score Sparsity

Table 3 presents the average number of unique relevance scores within the top-$k$ retrieved passages. While low-precision formats theoretically support a wide range of values, we observe a high density of scores in the $(0, 1)$ range, leading to frequent collisions. For instance, at $k = 100$, BF16 only yields 81.34 unique scores on average, indicating that nearly 20% of the documents share identical scores. The application of HPS noticeably restores score diversity even when using the same data type of model weights, bringing the counts closer to the FP32 baseline.

| | Average Number of Unique Scores at $k$ | | | | | | | |
|---|---|---|---|---|---|---|---|---|
| Data Type | 1 | 3 | 5 | 10 | 20 | 50 | 100 | 1K |
| BF16 | 1.00 | 2.83 | 4.67 | 9.22 | 18.18 | 43.37 | 81.34 | 443.81 |
| BF16 (+HPS) | 1.00 | 2.98 | 4.93 | 9.75 | 19.08 | 45.14 | 84.09 | 483.78 |
| FP16 | 1.00 | 2.95 | 4.89 | 9.78 | 19.48 | 48.31 | 95.11 | 712.73 |
| FP16 (+HPS) | 1.00 | 2.99 | 4.98 | 9.95 | 19.84 | 49.24 | 97.42 | 864.62 |
| FP32 | 1.00 | 3.00 | 4.99 | 9.99 | 19.97 | 49.92 | 99.85 | 998.69 |

Table 3: Average number of unique probabilities per query (↑). Results are averaged over the full MTEB-R datasets.

### C.2 Tie Group Size

Table 4 illustrates the average tie group size, defined as the ratio of $k$ to the number of unique scores. In BF16, we observe an average of 1.12 passages per tie group at $k = 10$. This phenomenon forces evaluators to rely on arbitrary tie-breaking heuristics (e.g., document IDs or metadata), which introduces noise into the ranking metrics. By reducing the tie group size and approaching the ideal of 1.00, HPS effectively mitigates this undesirable instability.

| | Average Tie Group Size at $k$ | | | | | | | |
|---|---|---|---|---|---|---|---|---|
| Data Type | 1 | 3 | 5 | 10 | 20 | 50 | 100 | 1K |
| BF16 | 1.0000 | 1.1073 | 1.1173 | 1.1241 | 1.1354 | 1.1795 | 1.2545 | 2.3840 |
| BF16 (+HPS) | 1.0000 | 1.0111 | 1.0172 | 1.0288 | 1.0515 | 1.1133 | 1.2008 | 2.2422 |
| FP16 | 1.0000 | 1.0284 | 1.0312 | 1.0305 | 1.0327 | 1.0405 | 1.0562 | 1.4973 |
| FP16 (+HPS) | 1.0000 | 1.0028 | 1.0048 | 1.0059 | 1.0089 | 1.0160 | 1.0270 | 1.1640 |
| FP32 | 1.0000 | 1.0018 | 1.0015 | 1.0016 | 1.0016 | 1.0016 | 1.0015 | 1.0013 |

Table 4: Average tie group size per query (↓). Lower values indicate fewer collisions.

### C.3 Rank Correlation against FP32

We verify whether the scores recovered by HPS closely align with the target order of the high-precision model. Table 5 shows the average Spearman's rank correlation ($\rho$) against the FP32 ranking, which we treat as the ground truth. The results confirm that HPS consistently improves the rank correlation. This demonstrates that HPS does not merely introduce random noise to break ties but successfully recovers meaningful semantic distinctions that are otherwise lost during low-precision computation.

| Data Type | Avg. $\rho$ |
|---|---|
| BF16 | 0.998960 |
| BF16 (+HPS) | 0.998961 |
| FP16 | 0.999972 |
| FP16 (+HPS) | 0.999975 |
| FP32 | 1.000000 |

Table 5: Average Spearman's rank correlation ($\rho$) against FP32 ground truth (↑).

## D Closed-form Expectations

Let the tie groups be $G_1, \dots, G_N$ in descending score order. Each group $G_n$ has size $|G_n|$ and $r_n$ relevant items ($0 \le r_n \le |G_n|$). Define the per-group relevance probability

$$p_n = \frac{r_n}{|G_n|}, \tag{6}$$

and the cumulative size

$$c_n = \sum_{m \le n} |G_m|, \qquad c_0 = 0. \tag{7}$$

For a cutoff rank $k$, the number of items from group $G_n$ that appear within the top-$k$ list is

$$t_n = \max\{0,\ \min(|G_n|,\ k - c_{n-1})\}. \tag{8}$$

**Count-based Metrics.**

With $N_+ = \sum_m r_m$,

$$\mathbb{E}[\text{Hits@}k] = \sum_{n : t_n > 0} p_n t_n, \tag{9}$$

$$\mathbb{E}[\text{Recall@}k] = \frac{\sum_n p_n t_n}{N_+}, \tag{10}$$

$$\mathbb{E}[\text{Precision@}k] = \frac{\sum_n p_n t_n}{k}, \tag{11}$$

$$\mathbb{E}[\text{F1@}k] = \frac{2 \sum_n p_n t_n}{k + N_+}. \tag{12}$$

**nDCG.**

With binary gains and weights $w_r = \frac{1}{\log_2(r+1)}$, define

$$W(a, b) = \sum_{r=a}^{b} w_r. \tag{13}$$

Then

$$\mathbb{E}[\mathrm{DCG}@k] = \sum_{n:t_n>0} p_n \, W(c_{n-1}+1,\ c_{n-1}+t_n), \tag{14}$$

$$\mathrm{IDCG}@k = \sum_{r=1}^{\min(N_+,k)} w_r, \tag{15}$$

$$\mathbb{E}[\mathrm{nDCG}@k] = \frac{\mathbb{E}[\mathrm{DCG}@k]}{\mathrm{IDCG}@k}. \tag{16}$$

**Reciprocal Rank.**

Let $n^* = \min\{n \mid r_n > 0\}$ be the first group containing a relevant item and $\binom{x_a}{x_b}$ be the binomial coefficient. If $k \le c_{n^*-1}$ then $\mathbb{E}[\mathrm{RR}@k] = 0$; otherwise

$$u = \min(|G_{n^*}| - 1,\ k - c_{n^*-1} - 1) \tag{17}$$

$$r_t = c_{n^*-1} + t + 1 \tag{18}$$

$$\pi_t = \frac{\binom{|G_{n^*}|-r_{n^*}}{t}}{\binom{|G_{n^*}|}{t}} \tag{19}$$

$$\lambda_t = \frac{r_{n^*}}{|G_{n^*}| - t} \tag{20}$$

$$\mathbb{E}[\mathrm{RR}@k] = \sum_{t=0}^{u} \frac{1}{r_t} \pi_t \lambda_t. \tag{21}$$

**Average Precision.**

For rank $r = c_{n-1} + t + 1$ with $0 \le t < t_n$ in group $G_n$,

$$A_{n,t} = R_{n-1} + 1 + t\frac{r_n - 1}{|G_n| - 1}, \tag{22}$$

$$D_{n,t} = c_{n-1} + t + 1, \tag{23}$$

where $R_{n-1} = \sum_{m<n} r_m$. The expected AP@$k$ is

$$\mathbb{E}[\mathrm{AP}@k] = \frac{1}{N_+} \sum_{n:r_n>0} \sum_{t=0}^{t_n-1} p_n \frac{A_{n,t}}{D_{n,t}}. \tag{24}$$

# E Time and Space Complexity

Let the ranked list for one query contain $L$ candidate documents and let the evaluation cutoff be $k$. The list is partitioned into $N$ tie groups $G_1, \dots, G_N$ of sizes $|G_1|, \dots, |G_N|$ with $\sum_{n=1}^{N} |G_n| = L$. All complexities below are per query.

**High-Precision Scoring (HPS).** Only the final logits are upcast to `FP32` and passed once through a scoring function $\phi$, so the time cost is $O(L)$ with negligible extra memory. In our implementation, converting 1,000,000 (batch size) × 1,024 (hidden size) `FP16` elements to `FP32` takes about 5 ms end-to-end on a single NVIDIA H200. From a memory standpoint, the impact is transient and bounded. If a temporary `FP32` buffer is materialized for the top-k block, the peak extra footprint is k × d × (4–2) bytes (e.g., 2 MB at k=1,024, d=1,024), and no `FP32` state is persisted after scoring.

**Tie-aware Retrieval Metric (TRM).** All computations occur after sorting, so no additional log $L$ factor is introduced. (i) A single left-to-right scan gathers the pairs $(|G_n|, r_n)$ for every tie group in $O(L)$ where $r_n$ refers to the number of relevant items in the $n$-th tie group $G_n$. (ii) Closed-form expressions let nDCG, MAP, Recall, Precision, and F1 be evaluated in $O(\min\{k, N\})$ time. (iii) MRR examines only the first tie group containing a relevant document, costing $O(|G_{j^*}|) \le O(k)$ where $j^*$ is the index of the tie group that includes the first relevant item. (iv) Max, min, and range scores need only the tie group that straddles rank $k$, again $O(k)$.

In total TRM adds at most $O(k + N) \subseteq O(L)$ lightweight arithmetic per query, far below the cost of the forward pass or initial sort, while providing tie-robust evaluation.

# F Extending to MTEB Retrieval

To verify that our findings generalize beyond the two primary datasets, we extend our evaluation to the MTEB-R benchmark. Our experiments on MTEB-R use 1K candidates retrieved by BM25.

Table 6 presents the quantitative stability results. We observe that standard `BF16` inference introduces high score range compared to the full precision (`FP32`) baseline. This significant metric instability is particularly pronounced in `Qwen3-Reranker-0.6B`, which exhibits the highest volatility with a range of 10.09 in nDCG@10 and 17.14 in MRR@10. In contrast, proposed HPS (`BF16`→`FP32`) effectively mitigates the instability, drastically reducing the range and absolute bias.

Figure 7 further visualizes this phenomenon for `Qwen3-Reranker-0.6B`. While the standard `BF16` regime (yellow) suffers from a wide variance in scores, our method (green) aligns closely with the ground truth `FP32` trajectory, ensuring reliable and consistent ranking evaluations.

## G Comparison with Alternative Baselines

In this section, we discuss why alternative tie-breaking strategies or numerical formulations are less ideal compared to the proposed protocol. We categorize these alternatives into stochastic, deterministic, numerical, and precision-based approaches.

### G.1 Stochastic Tie-breaking

A straightforward baseline to handle ties is to randomly permute the tied documents. However, this introduces non-determinism and sampling variance. Mathematically, if we denote the score of a random permutation baseline as $M_{\text{rand}}$, our *expected score* in TRM is equivalent to $\mathbb{E}[M_{\text{rand}}]$. While the empirical average of repeated random trials would converge to the expectation by the *Law of Large Numbers* (Ash and Doléans-Dade, 2000), it requires significant computational overhead to reduce variance. In contrast, TRM provides a closed-form, deterministic, and variance-free evaluation metric at negligible additional cost.

### G.2 Deterministic Heuristics

Deterministic tie-breaking policies often rely on the inherent storage order or external metadata (e.g., document IDs, timestamps). The *tie-oblivious* baseline ($M_{obl}$) discussed in Section 3 effectively corresponds to the former index-preserving policy.

A critical flaw in these heuristics is their susceptibility to spurious correlations. As shown in Figure 6, in datasets like AskUbuntuDupQuestions (Lei et al., 2016), positive passages are systematically indexed earlier than negative ones due to the data collection pipeline. Consequently, an index-preserving policy in this case acts identically to an optimistic $M_{\max}$ policy, leading to artificially inflated metrics. Relying on external metadata (e.g., chronological ordering) suffers from similar biases.

In contrast, HPS relies solely on model outputs, making it far less sensitive to metadata artifacts and thereby ensuring fairer and more reproducible scoring across varying storage implementations and data curation histories.

### G.3 Alternative Numerical Formulations

**Raw Logits vs. HPS (Sigmoid).** One might suggest using raw logits $z_i$ directly for ranking instead of upcasting followed by a sigmoid function, $\phi(\text{upcast}(z_i))$, arguing that logits offer a wider numerical range. However, we prove that distinct `FP16` logits map to distinct `FP32` sigmoid outputs, making the ranking identical.

Let $\sigma(x)$ be the sigmoid function. Its derivative is bounded by $0 < \sigma'(x) \leq 1/4$. By the *Mean Value Theorem*, for any two logits $z_i, z_j$, there exists $c$ such that:

$$|\sigma(z_i)-\sigma(z_j)| = \sigma'(c)|z_i-z_j| \leq \frac{1}{4}|z_i-z_j|. \quad (25)$$

For two `FP32` sigmoid outputs to collapse to the same value, their difference must be smaller than one `FP32` Unit in the Last Place (ULP) (Goldberg, 1991), which is approximately $10^{-7}$ in the $(0, 1)$ interval. Combining this with the inequality:

$$|z_i - z_j| \geq 4|\sigma(z_i) - \sigma(z_j)| \approx 4 \times 10^{-7}. \quad (26)$$

This implies that for ranking collisions to occur in HPS where they do not occur in raw logits, the input logits must differ by less than $\approx 4 \times 10^{-7}$. Since the precision grid of `BF16`/`FP16` is orders of magnitude coarser than this threshold (except in a negligible neighborhood around zero), **distinct 16-bit logits map to distinct `FP32` sigmoid scores**. Therefore, using raw logits provides no ranking benefit over HPS. Moreover, HPS preserves the $(0, 1)$ probability scale, maintaining interpretability and consistency across different models and datasets.

**Temperature Scaling.** Applying temperature scaling (Guo et al., 2017) to logits can mitigate saturation but does not solve the fundamental issue of the limited bucket count in 16-bit arithmetic. Furthermore, temperature is a hyperparameter that requires dataset-specific tuning. In contrast, HPS is a zero-parameter solution that universally mitigates spurious ties.

### G.4 Full Precision Computation

Performing all computations in `FP32` from the start is the ideal solution for numerical accuracy. However, this negates the efficiency benefits of modern low-precision accelerators (Micikevicius et al., 2017). As illustrated in Figure 7, proposed HPS yields performance trajectories that closely match the full `FP32` baseline, but with significantly lower memory bandwidth and computational costs as discussed in Appendix E. Thus, HPS offers the practically optimal trade-off between precision and efficiency.

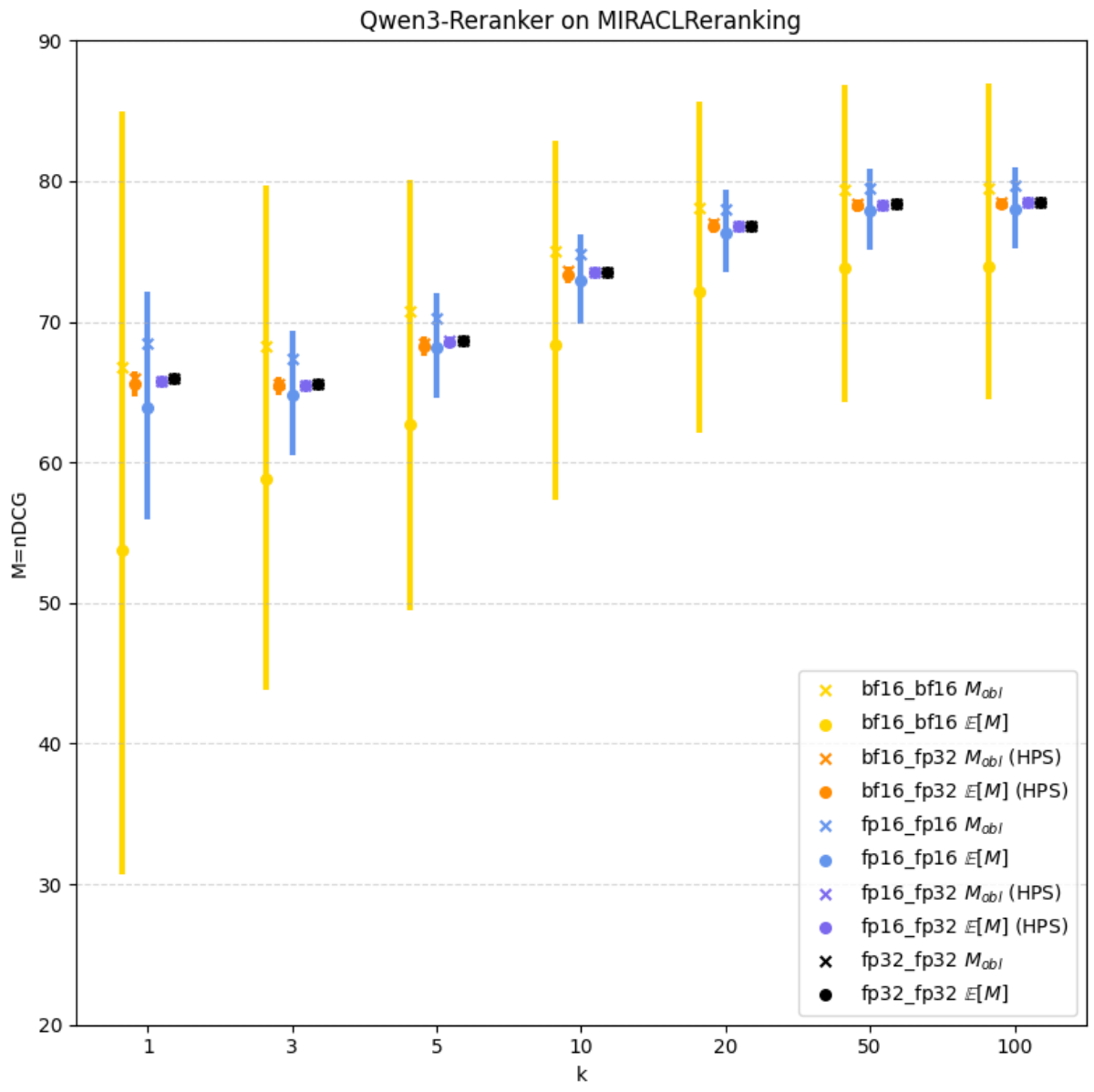

(a) nDCG

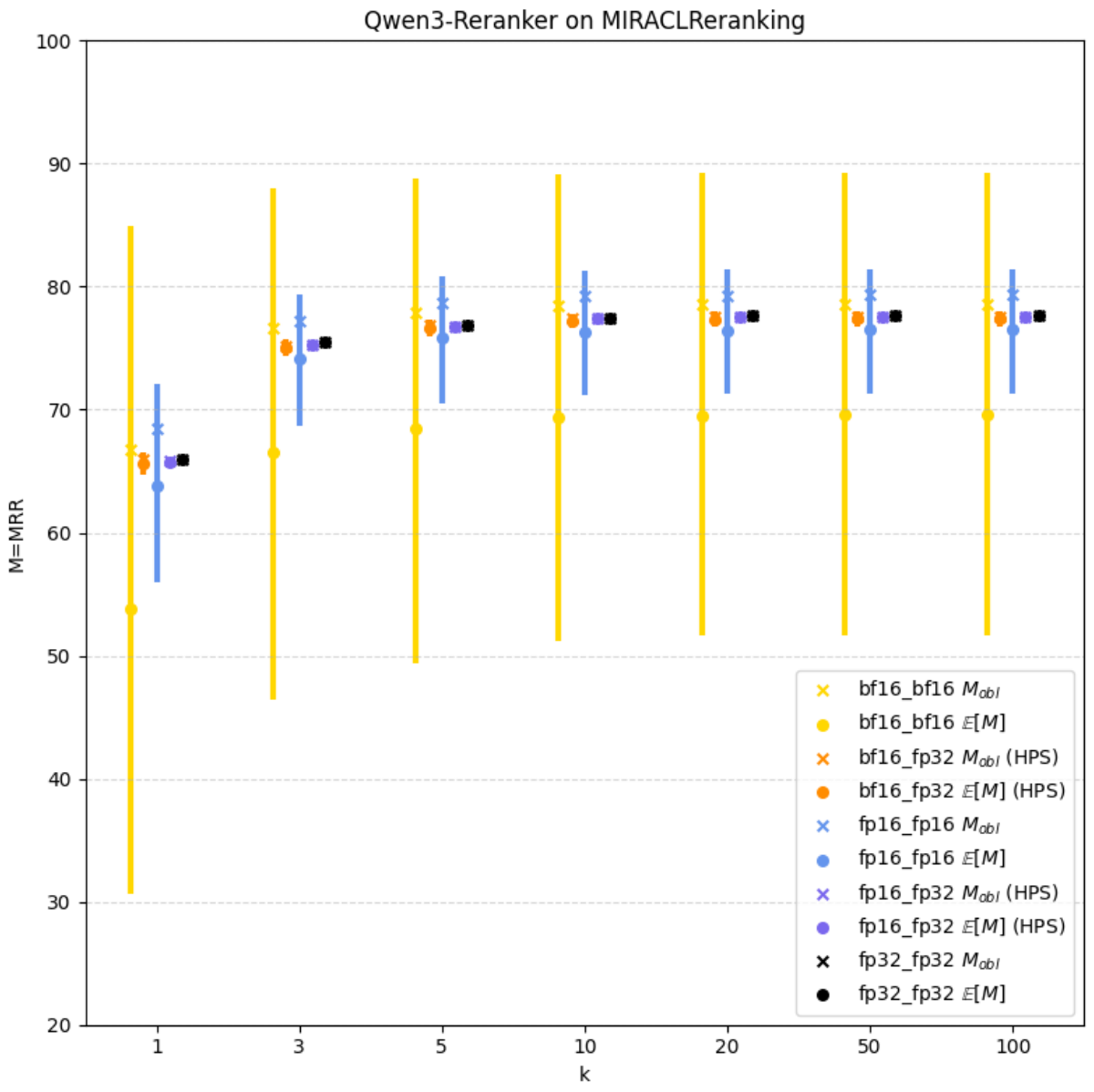

(b) MRR

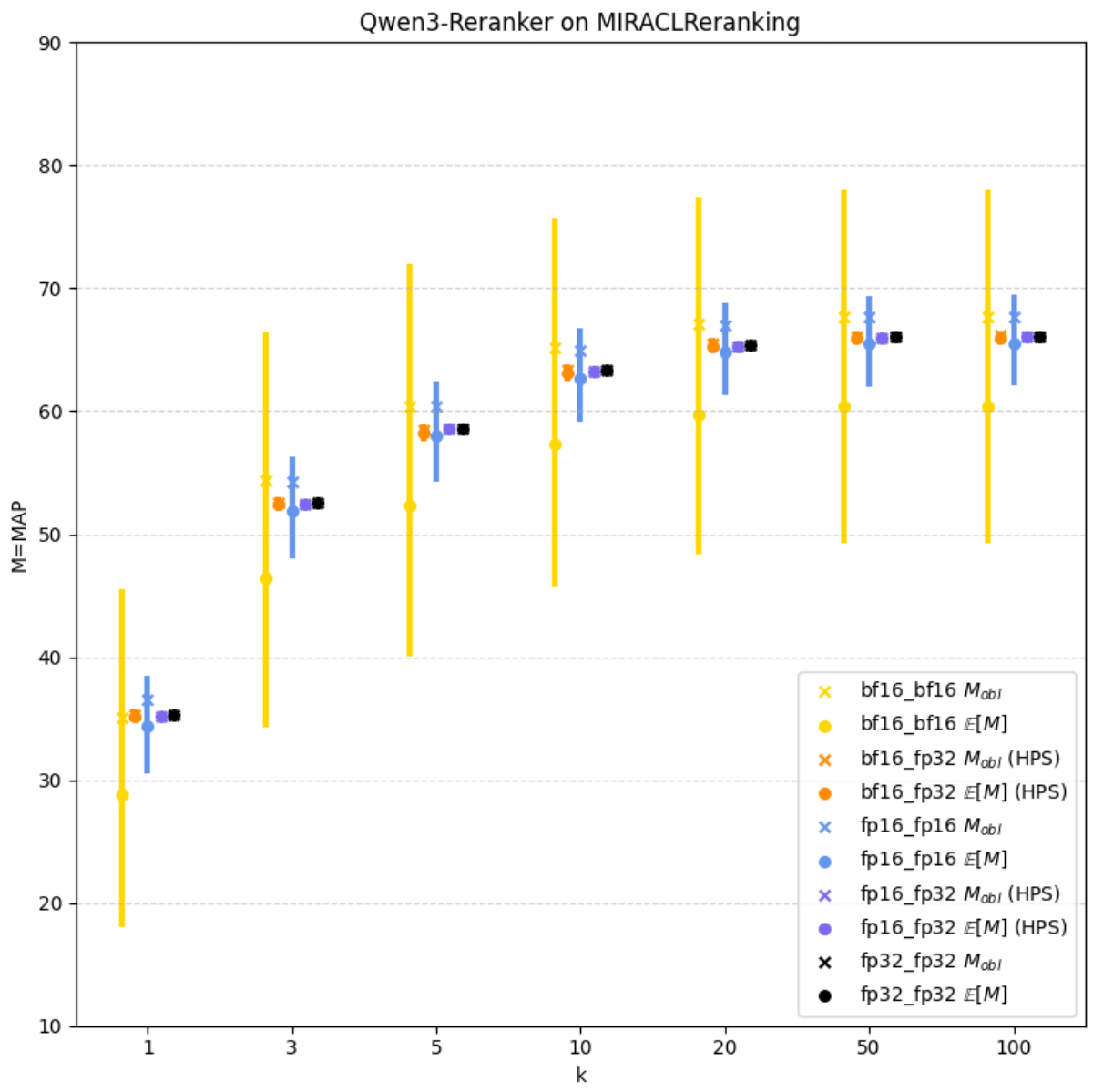

(c) MAP

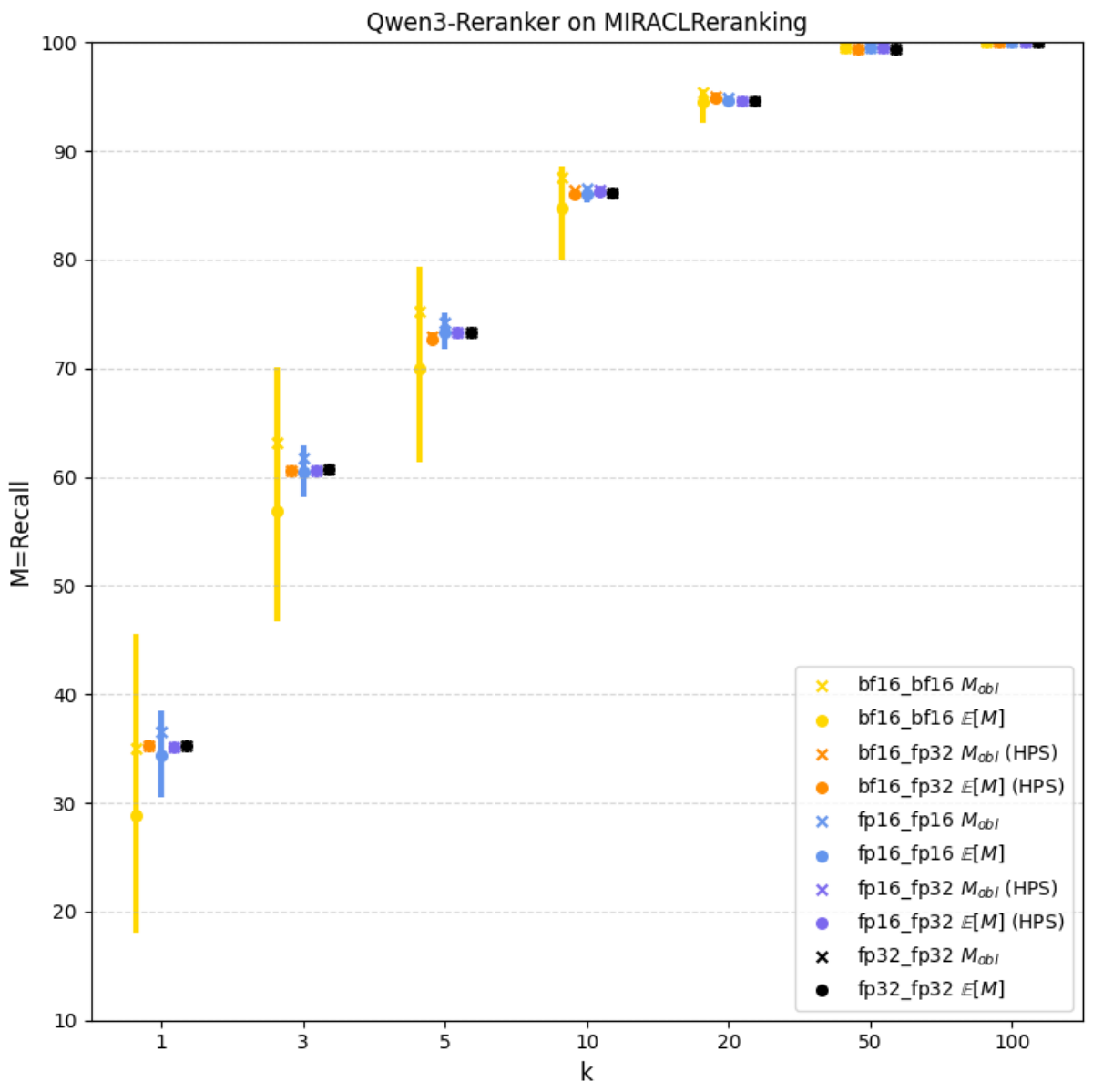

(d) Recall

Figure 5: Metric scores for cutoff $k$ of `Qwen3-Reranker-0.6B`, known as the state-of-the-are in text retrieval tasks, on **MIRACLReranking** dataset. Panels (a)-(d) report nDCG, MRR, MAP, and Recall. Each marker shows the tie-oblivious score $M_{obl}$ ($\times$) and the tie-aware expectation $\mathbb{E}[M]$ (⬤). The legend entry indicates the data types of the model and scoring function, respectively. For example, `BF16_FP32` denotes that the model operates in `BF16` precision, while the scoring function is upcast to `FP32`.

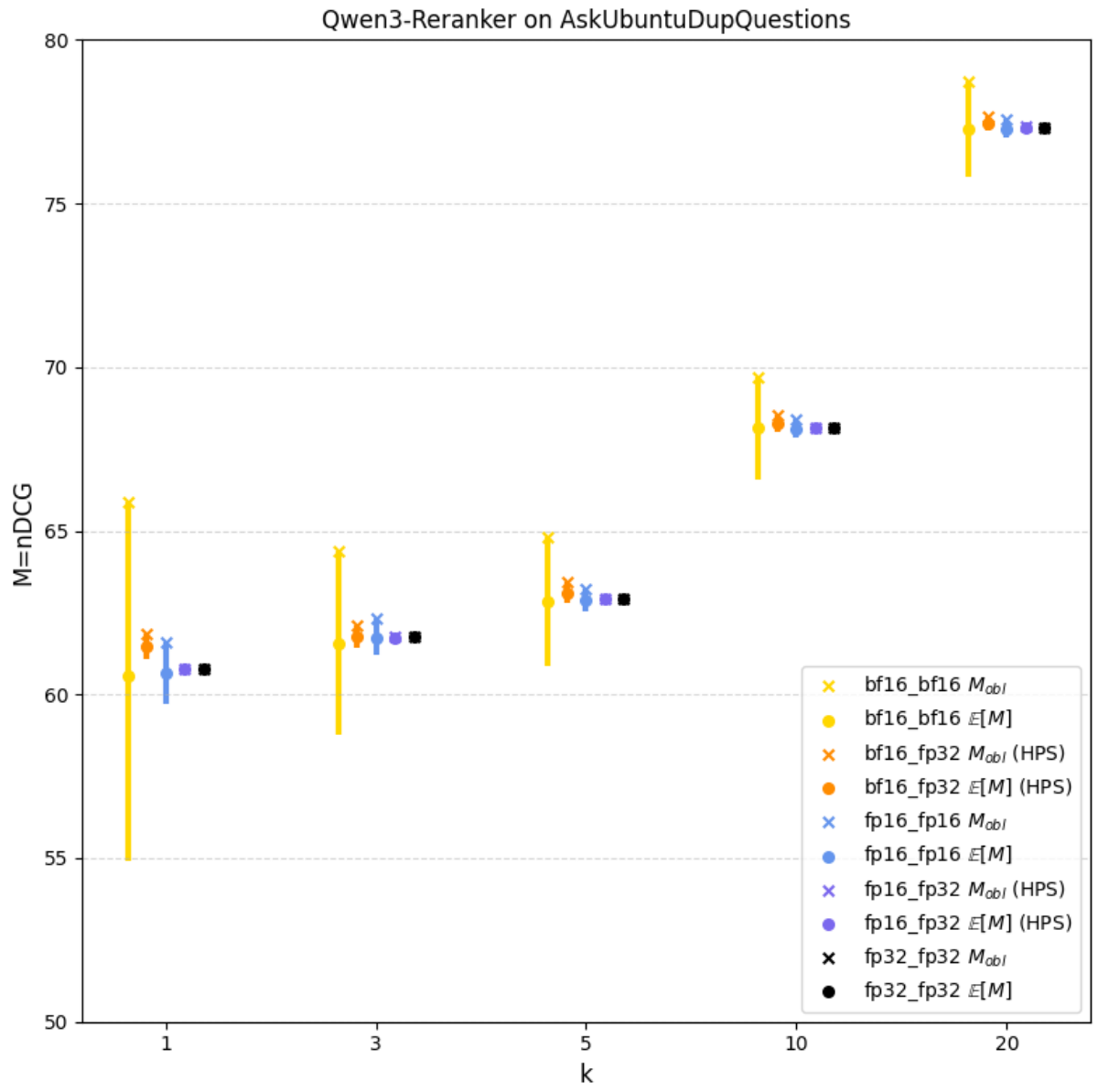

(a) nDCG

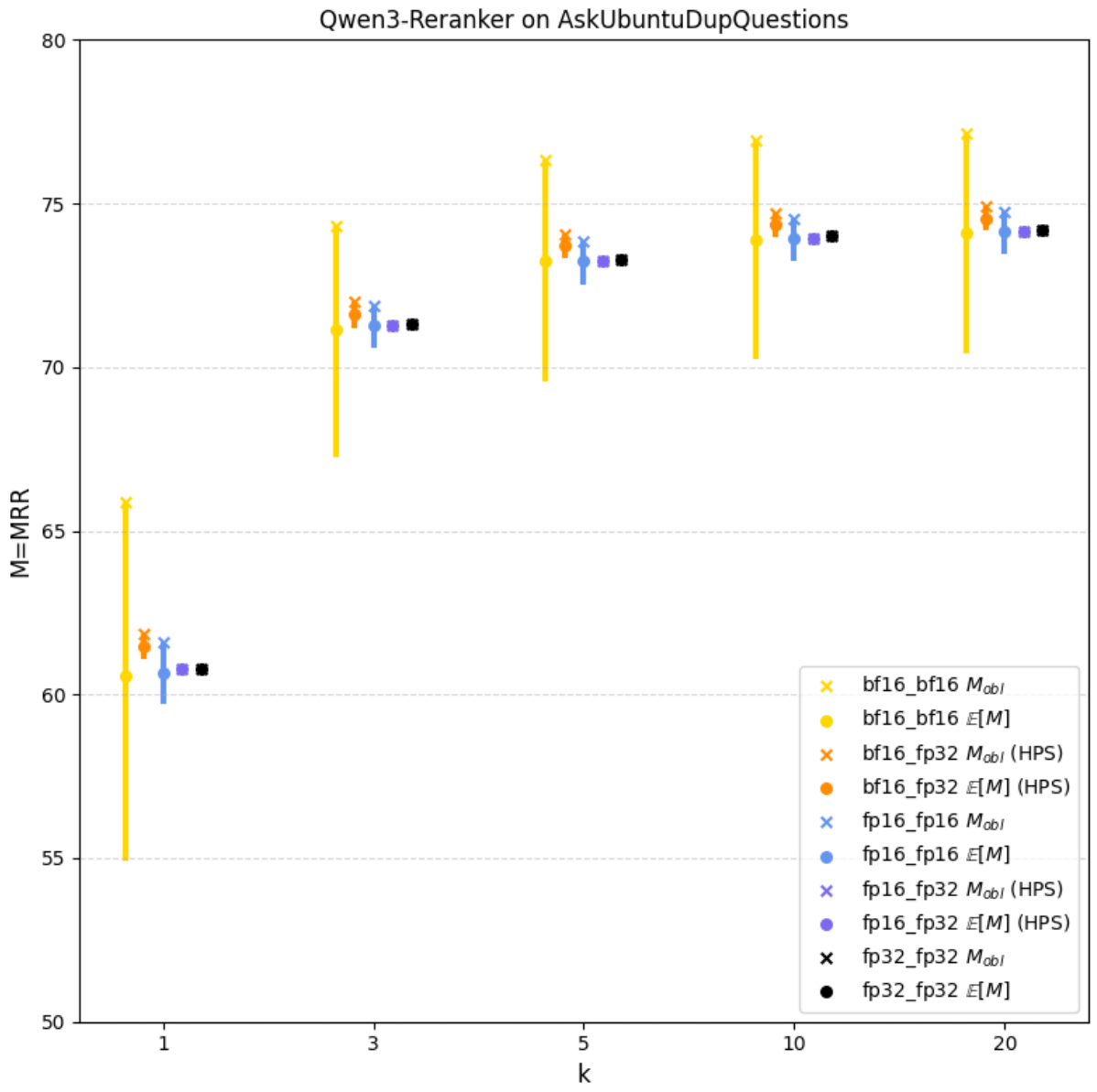

(b) MRR

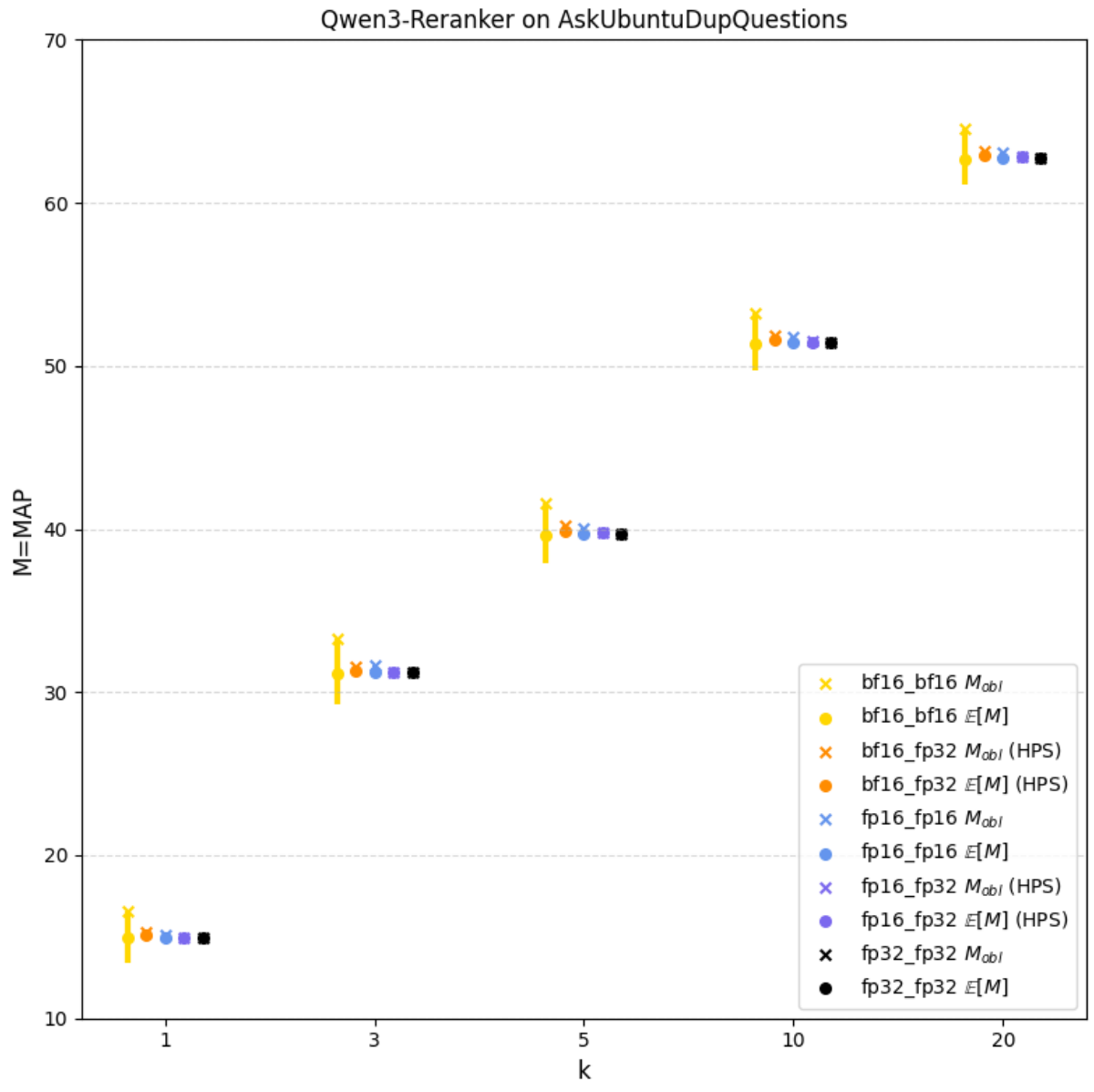

(c) MAP

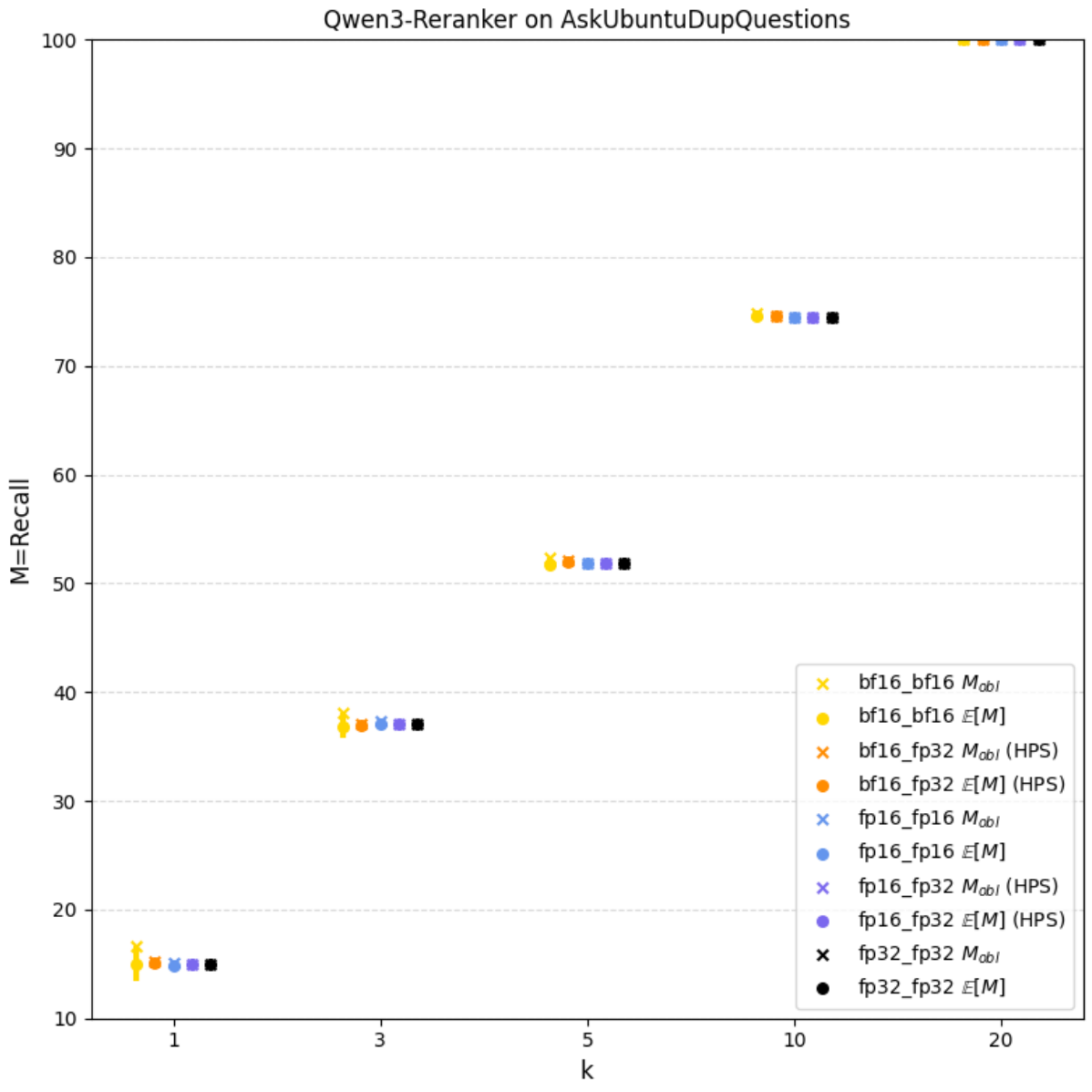

(d) Recall

Figure 6: Metric scores for cutoff $k$ of `Qwen3-Reranker-0.6B` on **AskUbuntuDupQuestions** dataset. In this dataset, all tie-oblivious metrics attain their maximum possible value (being overestimated) because, during candidate construction, every relevant item is concatenated ahead of all non-relevant ones.[6]

[6] https://github.com/embeddings-benchmark/mteb/blob/1.38.38/mteb/evaluation/evaluators/RerankingEvaluator.py#L175

| Models | FP32 | BF16 | | | | BF16 → FP32 (+HPS) | | | |
|---|---|---|---|---|---|---|---|---|---|
| | $M$ | $M_{obl}$ | $\mathbb{E}[M]$ | Range(▼) | \|Bias\|(▼) | $M_{obl}$ | $\mathbb{E}[M]$ | Range(▼) | \|Bias\|(▼) |
| ***M* = *nDCG*@10** | | | | | | | | | |
| `Qwen3-Reranker-0.6B`[♣] | 47.51 | 46.92 | 46.24 | 10.09 | 0.68 | 47.56 | 47.55 | **0.65** | **0.01** |
| `bge-reranker-v2-m3`[◇] | 43.94 | 43.70 | 43.81 | 2.13 | 0.11 | 43.96 | 43.92 | **0.18** | **0.03** |
| `gte-multilingual-reranker-base`[◇] | 46.72 | 46.71 | 46.62 | 0.93 | 0.09 | 46.70 | 46.69 | **0.12** | **0.01** |
| `Qwen3-Embedding-0.6B`[♠] | 45.59 | 45.55 | 45.50 | 1.19 | 0.06 | 45.50 | 45.50 | **0.00** | **0.00** |
| `multilingual-e5-large`[♠] | 43.20 | 43.66 | 43.13 | 4.46 | 0.53 | 43.35 | 43.35 | **0.00** | **0.00** |
| ***M* = *MRR*@10** | | | | | | | | | |
| `Qwen3-Reranker-0.6B`[♣] | 49.47 | 48.14 | 47.44 | 17.14 | 0.70 | 49.48 | 49.42 | **0.79** | **0.05** |
| `bge-reranker-v2-m3`[◇] | 45.38 | 44.92 | 45.34 | 4.27 | 0.42 | 45.36 | 45.35 | **0.17** | **0.00** |
| `gte-multilingual-reranker-base`[◇] | 49.42 | 49.11 | 49.14 | 1.41 | 0.03 | 49.37 | 49.36 | **0.14** | **0.01** |
| `Qwen3-Embedding-0.6B`[♠] | 47.23 | 47.36 | 47.22 | 1.47 | 0.14 | 47.31 | 47.31 | **0.00** | **0.00** |
| `multilingual-e5-large`[♠] | 44.57 | 45.65 | 44.83 | 5.60 | 0.82 | 44.95 | 44.95 | **0.00** | **0.00** |

Table 6: Evaluation stability results on the MTEB-R benchmark. All reported scores are averaged across ten diverse datasets, reranking top-1000 passages retrieved by BM25. We compare the tie-oblivious metric ($M_{obl}$), expectation ($\mathbb{E}[M]$), range (▼), and absolute bias (▼) under standard `BF16` and our `BF16`→`FP32` (+HPS). **Bold** values indicate the lowest range and absolute bias, and underlined values represent the highest performance in each metric. The application of HPS leads to a significant reduction in both range and absolute bias, effectively suppressing the numerical noise inherent in low-precision evaluation. Notably, while standard BF16 can lead to ranking inversions where the best-performing model in FP32 is no longer identified as the top choice, our HPS method restores the model rankings to be highly consistent with the full FP32 baseline.

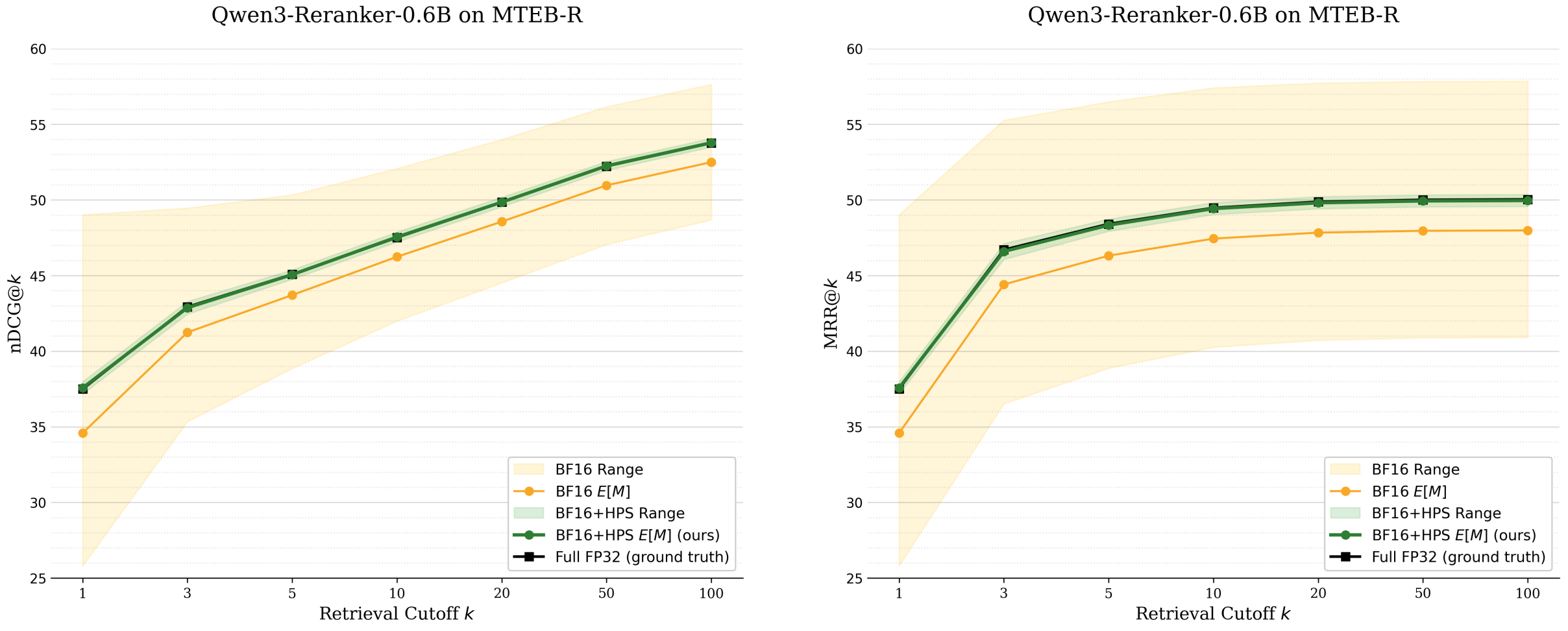

Figure 7: Visualization of **nDCG@*k* (left)** and **MRR@*k* (right)** fluctuations for `Qwen3-Reranker-0.6B` under the `BF16` precision regime on MTEB-R. The high variance in scores (yellow shaded area) demonstrates the inherent risk of reaching inconsistent ranking conclusions when using low-precision inference without a reliable protocol. In contrast, our proposed HPS yields performance trajectories that closely match the full FP32 baseline.